\documentclass[aps,prd,a4paper,11pt,onecolumn,groupedaddress,showpacs,nofootinbib,amssymb]{revtex4-2}
\usepackage[dvips]{graphicx}
\usepackage{amssymb}
\usepackage{amsmath}
\usepackage{mathrsfs}

\usepackage{graphicx,color}
\usepackage{amsfonts}
\usepackage{bm}
\usepackage{cancel}
\usepackage{comment}
\usepackage{slashed}
\usepackage{bbold}
\usepackage{physics}

\begin{document}

\title{Quasiclassical representation of the Volkov propagator and the tadpole diagram in a plane wave}
\author{A. Di Piazza}
\email{dipiazza@mpi-hd.mpg.de}
\affiliation{Max Planck Institute for Nuclear Physics, Saupfercheckweg 1, D-69117 Heidelberg, Germany}
\author{F. P. Fronimos}
\affiliation{Institute for Theoretical Physics, Heidelberg University, Philosophenweg 16, D-69120 Heidelberg, Germany}

\begin{abstract}
The solution of the Dirac equation in the presence of an arbitrary plane wave, corresponding to the so-called Volkov states, has provided an enormous insight in strong-field QED. In [Phys. Rev. A \textbf{103}, 076011 (2021)] a new ``fully quasiclassical'' representation of the Volkov states has been found, which is equivalent to the one known in the literature but which more transparently shows the quasiclassical nature of the quantum dynamics of an electron in a plane-wave field. Here, we derive the corresponding expression of the propagator by constructing it using the fully quasiclassical form of the Volkov states. The found expression allows one, together with the fully quasiclassical expression of the Volkov states, to compute probabilities in strong-field QED in an intense plane wave by manipulating only 2-by-2 rather than 4-by-4 Dirac matrices as in the usual approach. Moreover, apart from the exponential functions featuring the classical action of an electron in a plane wave, the fully quasiclassical Volkov propagator depends only on the electron kinetic four-momentum in the plane wave, which is a gauge-invariant quantity. Finally, we also compute the one-loop tadpole diagram in a plane wave starting from the Volkov propagator and we find that after renormalization it identically vanishes.
\end{abstract}

\maketitle

\section{Introduction}
The investigation of quantum electrodynamical processes occurring in the presence of intense laser fields has significantly improved our theoretical understanding of the strong-field regime of QED \cite{Mitter_1975,Ritus_1985,Ehlotzky_2009,Reiss_2009,Di_Piazza_2012,Roshchupkin_2012,
King_2016b,Blackburn_2020,Gonoskov_2021}. Strong-field QED signatures in the emission spectra of ultrarelativistic electrons colliding with an intense laser beam have also been observed in recent experiments \cite{Cole_2018,Poder_2018}. In the strong-field regime of QED the leptons involved in the processes experience in their rest frames field amplitudes of the order of or larger than the critical field of QED $F_{cr}=m^2/|e|$, where $m$ and $e<0$ are the electron mass and charge, respectively (units with $\hbar=c=\epsilon_0=1$ are used throughout). Moreover, the background field is so strong that it has to be taken into account exactly in the calculations by using the Furry picture \cite{Furry_1951,Landau_b_4_1982}. In the case of a background plane wave of electric-field amplitude $F_0$ and central angular frequency $\omega_0$, the latter condition corresponds to the so-called classical nonlinearity parameter $\xi_0=|e|F_0/m\omega_0$ being of the order of or larger than unity \cite{Mitter_1975,Ritus_1985,Ehlotzky_2009,Reiss_2009,Di_Piazza_2012}.

Analytical calculations within the Furry picture are feasible for background fields of sufficiently symmetric structure that the corresponding Dirac equation can be solved analytically \cite{Furry_1951,Landau_b_4_1982}. This is the case for a background plane-wave field, which is clearly relevant for processes occurring in the presence of background laser fields. The corresponding solutions of the Dirac equation are known as Volkov states \cite{Volkov_1935,Landau_b_4_1982} and have allowed for an enormous insight into strong-field QED in general and, in particular, into processes occurring in strong laser fields (see Refs. 
\cite{Reiss_1962,Nikishov_1964,Goldman_1964,Brown_1964,Boca_2009,Heinzl_2010b,
Mackenroth_2010,Mueller_2011b,Boca_2011,Mackenroth_2011,Seipt_2011,Seipt_2011b,
Dinu_2012,Titov_2012,Dinu_2013,Jansen_2013,Augustin_2014,Krajewska_2014,Seipt_2016,
Angioi_2016,Harvey_2016b,Di_Piazza_2018,Alexandrov_2019,Di_Piazza_2019,Ilderton_2019_b,
King_2020,Seipt_2020,King_2020_b} for the basic processes of nonlinear Compton scattering and nonlinear Breit-Wheeler pair production, Refs. \cite{Loetstedt_2009,Hu_2010,Ilderton_2011,Seipt_2012,Mackenroth_2013,King_2013, King_2015,Dinu_2018,Mackenroth_2018,Dinu_2019,Dinu_2020,Torgrimsson_2020,Bragin_2020,
Torgrimsson_2021} for higher-order processes, and Refs. \cite{Ritus_1970,Becker_1975,Baier_1976_a,Baier_1976_b,Narozhny_1979,Narozhny_1980,
Morozov_1981,Meuren_2013,Fedotov_2017,Podszus_2019,Ilderton_2019,Mironov_2020,
Di_Piazza_2020_b,Di_Piazza_2021_d} for radiative corrections).

It has been noticed that the Volkov states, although being an exact solution of the Dirac equation, have a quasiclassical form, in the sense that they feature the exponential of the classical action of an electron in a plane wave \cite{Ritus_1985}. Moreover, the spinoral structure of the Volkov states is such that the average spin four-vector satisfies the ``classical'' Bargmann-Michel-Telegdi equation in a plane wave. However, the spinorial structure of the Volkov states itself is not manifestly quasiclassical, i.e., it is not the same as that of spinors within the Wentzel-Kramers-Brillouin (WKB) approximation \cite{Pauli_1932,Rubinow_1963,Di_Piazza_2014}. In Ref. \cite{Di_Piazza_2021_a} one of us has found an alternative representation of the Volkov states, which is equivalent to the conventional one, but it is ``fully quasiclassical'' in the sense that its spinorial structure is also identical to that of WKB wave functions.

In the present paper, we continue this analysis, present an alternative derivation of the fully quasiclassical Volkov states, simpler than that in Ref. \cite{Di_Piazza_2021_a}, and compute the corresponding expression of the Volkov propagator, i.e., of the exact electron propagator in an arbitrary plane wave. Analogously to the fully quasiclassical Volkov states, the fully quasiclassical Volkov propagator is expressed as four blocks of 2-by-2 matrices, which involve the unity matrix and the Pauli matrices. This is reminiscent of the so-called ``spinor helicity formalism'', which is widely used in QCD in vacuum \cite{Dixon_2013}, as it simplifies some computations, and more recently in external plane waves \cite{Adamo_2019,Adamo_2020}. Moreover, the found expression of the Volkov propagator explicitly depends only on the dressed kinetic four-momentum of the electron in a plane wave, apart from exponential functions of the classical action, which transparently shows the transformation properties of the propagator under a generic gauge transformation of the background plane wave. Apart from its intrinsic interest, the fully quasiclassical Volkov propagator allows one, together with the fully quasiclassical Volkov states, to perform strong-field QED calculations only manipulating two-dimensional matrices/spinors rather than four-dimensional ones as conventionally done, and to directly obtain manifestly gauge-invariant results. An alternative and particularly simple representation of the Volkov propagator has been found in Ref. \cite{Lavelle_2021} in the case of a monochromatic, circularly polarized plane wave, by using a special gauge in which the plane-wave four-vector potential is orthogonal to the four-momentum entering the propagator.

Finally, in relation to the Volkov propagator, we investigate the tadpole diagram in a plane wave. This diagram in a constant background field has recently received attention \cite{Karbstein_2017,Edwards_2017,Ahmadiniaz_2017} since it was shown that the Euler-Heisenberg effective Lagrangian entails a two-loop one-particle reducible contribution \cite{Gies_2017}. It was found in Ref. \cite{Ahmadiniaz_2019} by using the worldline formalism that the contribution of the tadpole is linear in the external plane wave and that for this reason it can be renormalized out. Below, we show that for the plane-wave case the computation of the tadpole is significantly simplified by starting from the general definition of the vacuum four-current and from its relation with the Volkov propagator. In this way, we explicitly prove that the tadpole itself as well as its contribution to an arbitrary physical process are linear in the plane-wave field amplitude and that, after renormalization, it vanishes identically.

The paper is organized as follows. In Sect. II an alternative derivation of the fully quasiclassical Volkov states is obtained. In Sect. III the fully quasiclassical form of the Volkov propagator is obtained starting from the corresponding expression in terms of the fully quasiclassical Volkov states. In Sect. IV the tadpole diagram in a plane wave is investigated. In Sect. V the main conclusions of the paper are presented. Finally, an appendix contains technical details of a result presented in the main text.

Throughout this paper the Minkowski metric tensor is assumed to have the following signature $\eta^{\mu\nu}=\text{diag}(+1,-1,-1,-1)$ such that the Dirac gamma matrices $\gamma^{\mu}$ satisfy the anticommutation relation $\{\gamma^\mu,\gamma^\nu\}=2\eta^{\mu\nu}$ (the matrix $\gamma^5$ is defined as $\gamma^5=i\gamma^0\gamma^1\gamma^2\gamma^3$). In addition, the prime $'$ denotes the derivative with respect to the light-cone time $\phi=(nx)$, where $n^{\mu}=(1,\bm{n})$ and $n^2=0$. Finally, the hat notation on a four-vector stands for the contraction of the four-vector with the gamma matrices.

\section{Derivation of the fully quasiclassical form of the Volkov states}
The Volkov states are solutions of the Dirac equation in the presence of a plane-wave background field. For the sake of definiteness, we assume that the plane wave propagates along the $\bm{n}$ direction such that it can be described by a four-vector potential $A^{\mu}(\phi)$, which depends only on the quantity $\phi$ introduced above. By working in the Lorenz gauge $\partial_{\mu}A^{\mu}(\phi)=0$, with the additional conditions $A^0(\phi)=0$ and $\lim_{\phi\to\pm\infty}A^{\mu}(\phi)=0$, the four-vector $A^{\mu}(\phi)$ has the form $A^{\mu}(\phi)=(0,\bm{A}(\phi))$, with $\bm{n}\cdot\bm{A}(\phi)=0$.

We first recall that, since the vacuum is stable in a plane wave \cite{Schwinger_1951}, in- and out-states in a plane wave are physically equivalent. By limiting then to the in-states, the positive-energy Volkov states are conventionally written in the form \cite{Landau_b_4_1982}
\begin{equation}
\label{U}
U_{p,\sigma}(x)=e^{iS_p(x)}\left[1+e\frac{\hat{n}\hat{A}(\phi)}{2p_-}\right]u_{p,\sigma},
\end{equation}
where
\begin{equation}
\label{phase}
S_p(x)=-(px)-\int_{-\infty}^\phi{d\phi'\left[\frac{e(pA(\phi'))}{p_-}-\frac{e^2A^2(\phi')}{2p_-}\right]}
\end{equation}
is the classical action of an electron in the plane wave, $p^{\mu}=(\varepsilon_p,\bm{p})=(\sqrt{m^2+\bm{p}^2},\bm{p})$ is the asymptotic electron four-momentum for $\phi\to-\infty$ ($p_-=(np)$), and
\begin{equation}
\label{u}
u_{p,\sigma}=\begin{pmatrix}
\sqrt{\varepsilon_p+m}\xi_{p,\sigma}\\
\frac{\bm{p}\cdot\bm{\sigma}}{\sqrt{\varepsilon_p+m}}\xi_{p,\sigma}
\end{pmatrix}
\end{equation}
is the positive-energy free spinor characterized also by the spin quantum number $\sigma=\pm 1$ (we also assume a unity quantization volume). Here, the two-dimensional matrices $\bm{\sigma}$ are the Pauli matrices, whereas the two-dimensional spinor $\xi_{p,\sigma}$ describes the spin state of the electron and it is normalized as $\xi^{\dag}_{p,\sigma}\xi_{p,\sigma'}=\delta_{\sigma\sigma'}$ \cite{Landau_b_4_1982}.

In order to derive the fully quasiclassical form of the Volkov state $U_{p,\sigma}(x)$, we seek a solution $\Psi_{p,\sigma}(x)$ of the Dirac equation
\begin{equation}
\{\gamma^{\mu}[i\partial_{\mu}-eA_{\mu}(\phi)]-m\}\Psi_{p,\sigma}(x)=0
\end{equation}
in the plane-wave field $A^{\mu}(\phi)$ of the form
\begin{equation}
\label{Psi}
\Psi_{p,\sigma}(x)=e^{iS_p(x)}\Theta_{p,\sigma}(\phi).
\end{equation}
Hence, the spinor $\Theta_{p,\sigma}(\phi)$ has to satisfy the equation
\begin{equation}
\label{difeq1}
i\hat{n}\Theta'_{p,\sigma}(\phi)+[\hat\pi_p^{(e)}(\phi)-m]\Theta_{p,\sigma}(\phi)=0,
\end{equation}
where
\begin{equation}
\label{Pi}
\pi_p^{(e)\,\mu}(\phi)=(\varepsilon_p^{(e)}(\phi),\bm{\pi}_p^{(e)}(\phi))=-\partial^{\mu}S_p(x)-eA^{\mu}(\phi)=p^\mu-eA^\mu(\phi)+\frac{e(pA(\phi))}{p_-}n^{\mu}-\frac{e^2A^2(\phi)}{2p_-}n^{\mu}
\end{equation}
is the kinetic four-momentum of the electron in the plane wave with initial four-momentum $p^{\mu}$ at $\phi\to-\infty$. Note that Eq. (\ref{difeq1}) allows one to choose the spinor $\Theta_{p,\sigma}(\phi)$ to be only a function of $\phi$.
 
By recalling the general technique for solving the Dirac equation in an external field by passing to the ``quadratic'' Dirac equation \cite{Landau_b_4_1982}, we make the ansatz
\begin{equation}
\label{ansatz2}
\Theta_{p,\sigma}(\phi)=\frac{1}{2m}\left[\hat\pi_p^{(e)}(\phi)\Phi_{p,\sigma}(\phi)+m\Phi_{p,\sigma}(\phi)+i\hat{n}\Phi'_{p,\sigma}(\phi)\right],
\end{equation}
where $\Phi_{p,\sigma}(\phi)$ must satisfy the following equation
\begin{equation}
\label{difeq2}
2p_-\Phi'_{p,\sigma}(\phi)-e\hat{n}\hat A'(\phi)\Phi_{p,\sigma}(\phi)=0.
\end{equation}
Since $\Phi'_{p,\sigma}(\phi)$ has to be proportional to $\hat{n}$, we can conclude from Eq. (\ref{ansatz2}) that
\begin{equation}
\label{ansatz3}
\Theta_{p,\sigma}(\phi)=\frac{\hat\pi_p^{(e)}(\phi)+m}{2m}\Phi_{p,\sigma}(\phi).
\end{equation}
Now, by recalling the quasiclassical approach \cite{Pauli_1932,Rubinow_1963,Di_Piazza_2014}, we look for a solution, which also satisfies the ``vacuum-like'' equation \cite{Landau_b_4_1982}
\begin{equation}
\label{pi_eq}
[\hat{\pi}_p^{(e)}(\phi)-m]\Phi_{p,\sigma}(\phi)=0,
\end{equation}
which together with Eq. (\ref{ansatz3}) implies that $\Phi_{p,\sigma}(\phi)=\Theta_{p,\sigma}(\phi)$. From the solution of the Dirac equation in vacuum, it then follows that the spinor $\Theta_{p,\sigma}(\phi)$ has the form [see Eq. (\ref{u})]
\begin{equation}
\label{sol1}
\Theta_{p,\sigma}(\phi)=\begin{pmatrix} \sqrt{\varepsilon_p^{(e)}(\phi)+m}r^{(e)}_{p,\sigma}(\phi)\\ \frac{\bm\sigma\cdot\bm\pi_p^{(e)}(\phi)}{\sqrt{\varepsilon_p^{(e)}(\phi)+m}}r^{(e)}_{p,\sigma}(\phi)\end{pmatrix}=\frac{\hat\pi_p^{(e)}(\phi)+m}{\sqrt{\varepsilon_p^{(e)}(\phi)+m}}\begin{pmatrix} r^{(e)}_{p,\sigma}(\phi)\\0\end{pmatrix},
\end{equation}
where, as we will see, the two-dimensional spinor $r^{(e)}_{p,\sigma}(\phi)$ is related to the two-dimensional spinor $\xi_{p,\sigma}$ in Eq. (\ref{u}). One could think at this point that, since $[\hat{\pi}_p^{(e)}(\phi)-m]\Theta_{p,\sigma}(\phi)=0$, then Eq. (\ref{difeq1}) implies that $\Theta_{p,\sigma}(\phi)$ does not depend on $\phi$. However, Eq. (\ref{difeq1}) is a spinorial equation and one can see from Eq. (\ref{difeq2}) that the spinor $\hat{n}\Theta_{p,\sigma}(\phi)$ indeed does not depend on $\phi$ [recall that $\Phi_{p,\sigma}(\phi)=\Theta_{p,\sigma}(\phi)$].

Now, in order to determine the two-dimensional spinor $r^{(e)}_{p,\sigma}(\phi)$, it is convenient to introduce the electromagnetic field tensor $F^{\mu\nu}(\phi)=\partial^{\mu}A^{\nu}(\phi)-\partial^{\nu}A^{\mu}(\phi)=n^{\mu}A^{\prime\,\nu}(\phi)-n^{\nu}A^{\prime\,\mu}(\phi)$ of the plane wave by noticing that $\hat{n}\hat A'(\phi)=-(i/2)\sigma_{\mu\nu}F^{\mu\nu}(\phi)$, where $\sigma^{\mu\nu}=(i/2)[\gamma^{\mu},\gamma^{\nu}]$. By indicating as $\bm{E}(\phi)$ and $\bm{B}(\phi)$ the electric and magnetic field of the plane wave, respectively, we find that
\begin{equation}
\sigma_{\mu\nu}F^{\mu\nu}(\phi)=2i\bm{\alpha}\cdot\bm{E}(\phi)-2\bm{\Sigma}\cdot\bm{B}(\phi),
\end{equation}
where $\bm{\alpha}=\gamma^0\bm{\gamma}$ and $\bm{\Sigma}$ are the four-dimensional Pauli matrices, i.e., 
\begin{equation}
\bm{\Sigma}=\begin{pmatrix} \bm\sigma &\bm{0}\\
\bm{0} &\bm\sigma
\end{pmatrix}.
\end{equation}
In this way, Eq. (\ref{difeq2}) becomes
\begin{equation}
\label{difeq3}
2p_-\Phi'_{p,\sigma}(\phi)-e[\bm\alpha\cdot\bm E(\phi)+i\bm\Sigma\cdot\bm B(\phi)]\Phi_{p,\sigma}(\phi)=0,
\end{equation}
and it is satisfied by the ansatz in Eq. (\ref{sol1}) if the two-dimensional spinor $r^{(e)}_{p,\sigma}(\phi)$ satisfies the equation (see Ref. \cite{Di_Piazza_2021_a})
\begin{equation}
\label{difeq4}
r^{(e)\,\prime}_{p,\sigma}(\phi)=\frac{i e}{2p_-}\bm{\sigma}\cdot\left[\bm B(\phi)-\frac{\bm\pi_p^{(e)}(\phi)\times\bm E(\phi)}{\varepsilon_p^{(e)}(\phi)+m}\right]r^{(e)}_{p,\sigma}(\phi).
\end{equation}
In the Appendix A we prove explicitly that the solution of this differential equation with the initial condition $\lim_{\phi\to-\infty}r^{(e)}_{p,\sigma}(\phi)=\xi_{p,\sigma}$ is given by
\begin{equation}
\label{sol2}
\begin{split}
r^{(e)}_{p,\sigma}(\phi)&=\sqrt{\frac{\varepsilon_p^{(e)}(\phi)+m}{\varepsilon_p+m}}\left\{1-\frac{e}{2p_-}\bm\sigma\cdot\left[\bm{n}-\frac{\bm\pi_p^{(e)}(\phi)}{\varepsilon_p^{(e)}(\phi)+m}\right]\bm\sigma\cdot\bm A(\phi)\right\}\xi_{p,\sigma}\\
&=\sqrt{\frac{\varepsilon_p+m}{\varepsilon_p^{(e)}(\phi)+m}}\left[1+\frac{e}{2p_-}\bm\sigma\cdot\bm A(\phi)\bm\sigma\cdot\left(\bm{n}-\frac{\bm p}{\varepsilon_p+m}\right)\right]\xi_{p,\sigma}.
\end{split}
\end{equation}
Note that the prefactor $1/\sqrt{\varepsilon_p+m}$ is chosen in order for the normalization condition of the spinor to be $r_{p,\sigma}^{(e)\,\dagger} (\phi)r^{(e)}_{p,\sigma'}(\phi)=\xi_{p,\sigma}^\dagger \xi_{p,\sigma'}=\delta_{\sigma\sigma'}$. In this respect, the expression (\ref{sol2}) of the spinor $r^{(e)}_{p,\sigma}(\phi)$ can be written in a form, which transparently shows the conservation of the normalization of the spinor and which is also manifestly gauge invariant. To do this, we observe that
\begin{equation}
\begin{split}
e\bm A(\phi)&=\bm{p}-\bm\pi_p^{(e)}(\phi)-\bm{n}[\varepsilon_p-\varepsilon_p^{(e)}(\phi)]\\
&=\bm{n}[\varepsilon_p^{(e)}(\phi)+m]-\bm\pi_p^{(e)}(\phi)-\bm{n}(\varepsilon_p+m)+\bm{p}\\
&=[\varepsilon_p^{(e)}(\phi)+m]\left[\bm{n}-\frac{\bm\pi_p^{(e)}(\phi)}{\varepsilon_p^{(e)}(\phi)+m}\right]-(\varepsilon_p+m)\left(\bm{n}-\frac{\bm{p}}{\varepsilon_p+m}\right).
\end{split}
\end{equation}
In this way, we have that
\begin{equation}
\label{sol3}
\begin{split}
r^{(e)}_{p,\sigma}(\phi)&=\sqrt{\frac{\varepsilon_p^{(e)}(\phi)+m}{\varepsilon_p+m}}\left\{1-\frac{\varepsilon_p^{(e)}(\phi)+m}{2p_-}\left[\bm{n}-\frac{\bm\pi_p^{(e)}(\phi)}{\varepsilon_p^{(e)}(\phi)+m}\right]^2\right.\\
&\quad\left.+\frac{\varepsilon_p+m}{2p_-}\bm\sigma\cdot\left[\bm{n}-\frac{\bm\pi_p^{(e)}(\phi)}{\varepsilon_p^{(e)}(\phi)+m}\right]\bm\sigma\cdot\left(\bm{n}-\frac{\bm{p}}{\varepsilon_p+m}\right)\right\}\xi_{p,\sigma}\\
&=\frac{\sqrt{(\varepsilon_p+m)[\varepsilon_p^{(e)}(\phi)+m]}}{2p_-}\bm\sigma\cdot\left[\bm{n}-\frac{\bm\pi_p^{(e)}(\phi)}{\varepsilon_p^{(e)}(\phi)+m}\right]\bm\sigma\cdot\left(\bm{n}-\frac{\bm{p}}{\varepsilon_p+m}\right)\xi_{p,\sigma}.
\end{split}
\end{equation}
The conservation of the normalization is apparent because
\begin{align}
\left[\bm\sigma\cdot\left(\bm{n}-\frac{\bm{p}}{\varepsilon_p+m}\right)\right]^2&=\left(\bm{n}-\frac{\bm{p}}{\varepsilon_p+m}\right)^2=\frac{2p_-}{\varepsilon_p+m},\\
\left\{\bm{\sigma}\cdot\left[\bm{n}-\frac{\bm\pi_p^{(e)}(\phi)}{\varepsilon_p^{(e)}(\phi)+m}\right]\right\}^2&=\left[\bm{n}-\frac{\bm\pi_p^{(e)}(\phi)}{\varepsilon_p^{(e)}(\phi)+m}\right]^2=\frac{2p_-}{\varepsilon_p^{(e)}(\phi)+m}.
\end{align}

Finally, one can easily check that, as it should be, by substituting Eq. (\ref{sol2}) in Eq. (\ref{sol1}) and the resulting expression in Eq. (\ref{Psi}), one identically obtains that $\Psi_{p,\sigma}(x)=U_{p,\sigma}(x)$ and then that
\begin{equation}
\label{U_QC}
U_{p,\sigma}(x)=e^{iS_p(x)}\begin{pmatrix} \sqrt{\varepsilon_p^{(e)}(\phi)+m}r^{(e)}_{p,\sigma}(\phi)\\ \frac{\bm\sigma\cdot\bm\pi_p^{(e)}(\phi)}{\sqrt{\varepsilon_p^{(e)}(\phi)+m}}r^{(e)}_{p,\sigma}(\phi)\end{pmatrix}.
\end{equation}
Hence, apart from the action in the exponential\footnote{Note that in the chosen gauge with $A^0(\phi)=0$ and then $\bm{n}\cdot\bm{A}(\phi)=0$, the action $S_p(x)$ can be written as
\begin{equation}
\begin{split}
S_p(x)&=-p_-x_++\bm{p}_{\perp}\cdot\bm{x}_{\perp}-\frac{m^2+\bm{p}_{\perp}^2}{2p_-}\phi-\int_0^\phi d\phi'\left[-\frac{e\bm{p}_{\perp}\cdot\bm{A}_{\perp}(\phi')}{p_-}+\frac{e^2\bm{A}_{\perp}^2(\phi')}{2p_-}\right]+C\\
&=-p_-x_++\bm{p}_{\perp}\cdot\bm{x}_{\perp}-\int_0^\phi d\phi'\frac{m^2+\bm{\pi}^{(e)\,2}_{p,\perp}(\phi')}{2p_-}+C,
\end{split}
\end{equation}
where $x_+=(t+\bm{n}\cdot\bm{x})/2$, $\bm{x}_{\perp}=\bm{x}-(\bm{n}\cdot\bm{x})\bm{n}$ [analogous definitions hold for $\bm{p}_{\perp}$, $\bm{A}_{\perp}(\phi)=\bm{A}(\phi)$, and $\bm{\pi}^{(e)}_{p,\perp}(\phi)$], and $C=-\int_{-\infty}^0d\phi'\left[2e(pA(\phi'))-e^2A^2(\phi')\right]/2p_-$ is a physically irrelevant constant. However, this does not imply that Volkov states are, even apart from the constant $C$, gauge invariant (which would be incorrect, see, e.g., \cite{Landau_b_4_1982}), because the above expression of the action is valid only in the chosen gauge.}, the Volkov state $U_{p,\sigma}(x)$ can effectively be expressed only in terms of the initial four-momentum $p^{\mu}$ and of the electron (kinetic) four-momentum in the plane wave $\pi^{(e)\,\mu}_p(\phi)$.

From the fact that the Volkov state $U_{p,\sigma}(x)$ satisfies the equation $[\hat{\pi}_p^{(e)}(\phi)-m]U_{p,\sigma}(x)=0$ [see Eq. (\ref{pi_eq})] one could conclude that it can be written as the free state $u_{p,\sigma}$ [see Eq. (\ref{u})] with the electron four-momentum being replaced by the electron kinetic four-momentum in the plane wave \cite{Ilderton_2020_b}. However, this substitution rule does not apply to the two-dimensional spinors $\xi_{p,\sigma}$ in the free state in Eq. (\ref{u}) and $r_{p,\sigma}^{(e)}(\phi)$ in the Volkov state $U_{p,\sigma}(x)$ in Eq. (\ref{U_QC}). The former two-dimensional spinor is arbitrary in the free state, whereas the two-dimensional spinor $r_{p,\sigma}^{(e)}(\phi)$ has a determined, non-trivial time evolution [see Eq. (\ref{sol2})] as it has to satisfy Eq. (\ref{difeq4}) (see also the Appendix A) and only its initial condition is arbitrary. One can gain a more clear insight on the above argument by observing that
\begin{equation}
\label{p_pi}
\left[1+e\frac{\hat{n}\hat{A}(\phi)}{2p_-}\right]\hat{p}=\hat{\pi}^{(e)}_p(\phi)\left[1+e\frac{\hat{n}\hat{A}(\phi)}{2p_-}\right].
\end{equation}
This identity allows one to write [see Eq. (\ref{u})]
\begin{equation}
\begin{split}
\left[1+e\frac{\hat{n}\hat{A}(\phi)}{2p_-}\right]u_p&=\left[1+e\frac{\hat{n}\hat{A}(\phi)}{2p_-}\right]\frac{\hat{p}+m}{\sqrt{\varepsilon_p+m}}\begin{pmatrix}
\xi_{p,\sigma}\\
0
\end{pmatrix}=\frac{\hat{\pi}^{(e)}_p(\phi)+m}{\sqrt{\varepsilon_p+m}}\left[1+e\frac{\hat{n}\hat{A}(\phi)}{2p_-}\right]\begin{pmatrix}
\xi_{p,\sigma}\\
0
\end{pmatrix}\\
&=\frac{\hat{\pi}^{(e)}_p(\phi)+m}{\sqrt{\varepsilon_p+m}}\begin{pmatrix}
\left[1-\frac{e}{2p_-}\bm{\sigma}\cdot\bm{n}\bm{\sigma}\cdot\bm{A}(\phi)\right]\xi_{p,\sigma}\\
-\frac{e}{2p_-}\bm{\sigma}\cdot\bm{A}(\phi)\xi_{p,\sigma}
\end{pmatrix}.
\end{split}
\end{equation}
This equation shows that the spinor to which the matrix $\hat{\pi}^{(e)}_p(\phi)+m$ is applied does not apparently feature a vanishing lower two-dimensional spinor as in Eq. (\ref{sol1}). However, the equivalence with Eq. (\ref{sol1}) is obtained by noticing that
\begin{equation}
\begin{split}
&\frac{\hat\pi_p^{(e)}(\phi)+m}{\sqrt{\varepsilon_p^{(e)}(\phi)+m}}\begin{pmatrix} r^{(e)}_{p,\sigma}(\phi)\\0\end{pmatrix}=\frac{\hat\pi_p^{(e)}(\phi)+m}{\sqrt{\varepsilon_p+m}}\begin{pmatrix} \left[1-\frac{e}{2p_-}\bm{\sigma}\cdot\bm{n}\bm{\sigma}\cdot\bm{A}(\phi)+\frac{e}{2p_-}\frac{\bm{\sigma}\cdot\bm{\pi}^{(e)}_p(\phi)\bm{\sigma}\cdot\bm{A}(\phi)}{\varepsilon_p^{(e)}(\phi)+m}\right]\xi_{p,\sigma}\\0\end{pmatrix}\\
&\quad=\frac{\hat\pi_p^{(e)}(\phi)+m}{\sqrt{\varepsilon_p+m}}\left\{\begin{pmatrix} \left[1-\frac{e}{2p_-}\bm{\sigma}\cdot\bm{n}\bm{\sigma}\cdot\bm{A}(\phi)\right]\xi_{p,\sigma}\\-\frac{e}{2p_-}\bm{\sigma}\cdot\bm{A}(\phi)\xi_{p,\sigma}\end{pmatrix}+\begin{pmatrix} \frac{e}{2p_-}\frac{\bm{\sigma}\cdot\bm{\pi}^{(e)}_p(\phi)\bm{\sigma}\cdot\bm{A}(\phi)}{\varepsilon_p^{(e)}(\phi)+m}\xi_{p,\sigma}\\\frac{e}{2p_-}\bm{\sigma}\cdot\bm{A}(\phi)\xi_{p,\sigma}\end{pmatrix}\right\}
\end{split}
\end{equation}
and that the second spinor belongs to the null space of the matrix $\hat{\pi}^{(e)}_p(\phi)+m$, which is the non-trivial point here.

It is known that the average four-momentum $P_p^{(e)\,\mu}(\phi)=m\bar{U}_{p,\sigma}(x)\gamma^{\mu}U_{p,\sigma}(x)/\bar{U}_{p,\sigma}(x)U_{p,\sigma}(x)$ (for a generic spinor $\psi$, it is $\bar{\psi}=\psi^{\dag}\gamma^0$) and the average spin four-vector\\ $S_{p,\sigma}^{(e)\,\mu}(\phi)=-\bar{U}_{p,\sigma}(x)\gamma^5\gamma^{\mu}U_{p,\sigma}(x)/\bar{U}_{p,\sigma}(x)U_{p,\sigma}(x)$ of an electron in a plane wave are given by (see, e.g., Ref. \cite{Ritus_1985})
\begin{align}
P_p^{(e)\,\mu}(\phi)&=\pi_p^{(e)\,\mu}(\phi),\\
S_{p,\sigma}^{(e)\,\mu}(\phi)&=s_{p,\sigma}^{\mu}-eA^{\mu}(\phi)\frac{s_{p,\sigma,-}}{p_-}+e(s_{p,\sigma}A(\phi))\frac{n^{\mu}}{p_-}-e^2A^2(\phi)\frac{s_{p,\sigma,-}}{2p^2_-}n^{\mu},
\end{align}
where $s_{p,\sigma}^{\mu}=-\bar{u}_{p,\sigma}\gamma^5\gamma^{\mu}u_{p,\sigma}/\bar{u}_{p,\sigma}u_{p,\sigma}$ is the initial average spin four-vector. It is already clear from their definitions that the four-vectors $P_p^{(e)\,\mu}(\phi)$ and $S_{p,\sigma}^{(e)\,\mu}(\phi)$ only depend on $\phi$ [also one sees that, as expected, the average four-momentum $P_p^{(e)\,\mu}(\phi)$ does not depend on the spin quantum number].

Now, from the structure of the state $U_{p,\sigma}(x)$ we can conclude that the four-vector $S_p^{(e)\,\mu}(\phi)$ must have the form (see Ref. \cite{Landau_b_4_1982} for the corresponding equation in vacuum)
\begin{equation}
S_{p,\sigma}^{(e)\,\mu}(\phi)=\left(\frac{\bm{S}^{(e)}_{p,\sigma,0}(\phi)\cdot\bm{\pi}_p^{(e)}(\phi)}{m},\bm{S}^{(e)}_{p,\sigma,0}(\phi)+\frac{\bm{S}^{(e)}_{p,\sigma,0}(\phi)\cdot\bm{\pi}_p^{(e)}(\phi)}{m[\varepsilon_p^{(e)}(\phi)+m]}\bm{\pi}_p^{(e)}(\phi)\right),
\end{equation}
where $\bm{S}^{(e)}_{p,\sigma,0}(\phi)=r_{p,\sigma}^{(e)\,\dagger}(\phi)\bm{\sigma}r^{(e)}_{p,\sigma}(\phi)$ corresponds to the three-dimensional spin vector in the instantaneous rest frame of the electron in the plane wave [where $\bm{\pi}_p^{(e)}(\phi)=\bm{0}$]. The vector $\bm{S}^{(e)}_{p,\sigma,0}(\phi)$ can be calculated explicitly by using Eq. (\ref{sol3}) and the result is
\begin{equation}
\begin{split}
\bm{S}^{(e)}_{p,\sigma,0}(\phi)&=\bm{s}^{(e)}_{p,\sigma,0}+\frac{\varepsilon_p+m}{p_-}\left(\bm{n}-\frac{\bm{p}}{\varepsilon_p+m}\right)\cdot\bm{s}^{(e)}_{p,\sigma,0}\left(\bm{n}-\frac{\bm{p}}{\varepsilon_p+m}\right)\\
&\quad-\frac{\varepsilon_p^{(e)}(\phi)+m}{p_-}\left[\bm{n}-\frac{\bm\pi_p^{(e)}(\phi)}{\varepsilon_p^{(e)}(\phi)+m}\right]\cdot\bm{s}^{(e)}_{p,\sigma,0}\left[\bm{n}-\frac{\bm\pi_p^{(e)}(\phi)}{\varepsilon_p^{(e)}(\phi)+m}\right]\\
&\quad+\frac{(\varepsilon_p+m)[\varepsilon_p^{(e)}(\phi)+m]}{p^2_-}\left(\bm{n}-\frac{\bm{p}}{\varepsilon_p+m}\right)\cdot\bm{s}^{(e)}_{p,\sigma,0}\\
&\qquad\times\left[\bm{n}-\frac{\bm\pi_p^{(e)}(\phi)}{\varepsilon_p^{(e)}(\phi)+m}\right]\times\left\{\left[\bm{n}-\frac{\bm\pi_p^{(e)}(\phi)}{\varepsilon_p^{(e)}(\phi)+m}\right]\times\left(\bm{n}-\frac{\bm{p}}{\varepsilon_p+m}\right)\right\},
\end{split}
\end{equation}
where $\bm{s}^{(e)}_{p,\sigma,0}=\xi_{p,\sigma}^\dagger \bm{\sigma}\xi_{p,\sigma}=\sigma\bm{s}_p$, with $\bm{s}_p=\xi_{p,+}^\dagger \bm{\sigma}\xi_{p,+}$ (this is equivalent to choosing the spinor $\xi_{p,\sigma}$ such that $\bm{\sigma}\cdot\bm{s}_p\xi_{p,\sigma}=\sigma\xi_{p,\sigma}$). Note that the quantity $\bm{S}^{(e)}_{p,\sigma,0}(\phi)$ can also be obtained from the expression of $S_{p,\sigma}^{(e)\,\mu}(\phi)=(S_{p,\sigma}^{(e)\,0}(\phi),\bm{S}_{p,\sigma}^{(e)}(\phi))$ as $\bm{S}^{(e)}_{p,\sigma,0}(\phi)=\bm{S}_{p,\sigma}^{(e)}(\phi)-S_{p,\sigma}^{(e)\,0}(\phi)\bm\pi_p^{(e)}(\phi)/[\varepsilon_p^{(e)}(\phi)+m]$.

The case of negative-energy states $V_{p,\sigma}(x)$ can be worked out analogously. One starts with the same ansatz as in Eq. (\ref{Psi}) but with the action $S_p(x)$ being replaced by $S_{-p}(x)$. Then, one chooses the solution in such a way that it resembles the free negative-energy spinor
\begin{equation}
\label{v}
v_{p,\sigma}=\begin{pmatrix}
\frac{\bm{p}\cdot\bm{\sigma}}{\sqrt{\varepsilon_p+m}}\chi_{p,\sigma}\\
\sqrt{\varepsilon_p+m}\chi_{p,\sigma}
\end{pmatrix},
\end{equation}
where the two-dimensional spinor $\chi_{p,\sigma}$ can be chosen as $\chi_{p,\sigma}=-i\sigma_2\xi^*_{p,\sigma}$ and it is then normalized as $\chi^{\dag}_{p,\sigma}\chi_{p,\sigma'}=\delta_{\sigma\sigma'}$ \cite{Landau_b_4_1982}. This is achieved by requiring that $V_{p,\sigma}(x)$ satisfies the equation $[\hat{\pi}_p^{(p)}(\phi)+m]V_{p,\sigma}(x)=0$, where
\begin{equation}
\label{Pi_p}
\pi_p^{(p)\,\mu}(\phi)=(\varepsilon_p^{(p)}(\phi),\bm{\pi}_p^{(p)}(\phi))=\partial^{\mu}S_{-p}(x)+eA^{\mu}(\phi)=p^\mu+eA^\mu(\phi)-\frac{e(pA(\phi))}{p_-}n^{\mu}-\frac{e^2A^2(\phi)}{2p_-}n^{\mu}
\end{equation}
is the classical kinetic four-momentum of a positron in the plane wave with initial four-momentum $p^{\mu}$ at $\phi\to-\infty$.

By following the same steps as in the positive-energy case, we obtain
\begin{equation}
\label{V_QC}
V_{p,\sigma}(x)=e^{iS_{-p}(x)}\begin{pmatrix}\frac{\bm\sigma\cdot\bm\pi_p^{(p)}(\phi)}{\sqrt{\varepsilon_p^{(p)}(\phi)+m}}r^{(p)}_{p,\sigma}(\phi) \\ \sqrt{\varepsilon_p^{(p)}(\phi)+m}r^{(p)}_{p,\sigma}(\phi)\end{pmatrix},
\end{equation}
where
\begin{equation}
\begin{split}
r^{(p)}_{p,\sigma}(\phi)&=\sqrt{\frac{\varepsilon_p^{(p)}(\phi)+m}{\varepsilon_p+m}}\left\{1+\frac{e}{2p_-}\bm\sigma\cdot\left[\bm{n}-\frac{\bm\pi_p^{(p)}(\phi)}{\varepsilon_p^{(p)}(\phi)+m}\right]\bm\sigma\cdot\bm A(\phi)\right\}\chi_{p,\sigma}\\
&=\sqrt{\frac{\varepsilon_p+m}{\varepsilon_p^{(p)}(\phi)+m}}\left[1-\frac{e}{2p_-}\bm\sigma\cdot\bm A(\phi)\bm\sigma\cdot\left(\bm{n}-\frac{\bm p}{\varepsilon_p+m}\right)\right]\chi_{p,\sigma}.
\end{split}
\end{equation}
Also, it can be shown that the fully quasiclassical spinor in Eq. (\ref{V_QC}) can be also written in the conventional form as \cite{Landau_b_4_1982}
\begin{equation}
V_{p,\sigma}(x)=e^{iS_{-p}(x)}\left[1-e\frac{\hat{n}\hat{A}(\phi)}{2p_-}\right]v_{p,\sigma}.
\end{equation}
Finally, by using the identity
\begin{equation}
e\bm A(\phi)=(\varepsilon_p+m)\left(\bm{n}-\frac{\bm{p}}{\varepsilon_p+m}\right)-[\varepsilon_p^{(p)}(\phi)+m]\left[\bm{n}-\frac{\bm\pi_p^{(p)}(\phi)}{\varepsilon_p^{(p)}(\phi)+m}\right],
\end{equation}
we can write the two-dimensional spinor $r^{(p)}_{p,\sigma}(\phi)$ in the manifestly gauge-invariant form as
\begin{equation}
\label{sol5}
r^{(p)}_{p,\sigma}(\phi)=\frac{\sqrt{(\varepsilon_p+m)[\varepsilon_p^{(p)}(\phi)+m]}}{2p_-}\bm\sigma\cdot\left[\bm{n}-\frac{\bm\pi_p^{(p)}(\phi)}{\varepsilon_p^{(p)}(\phi)+m}\right]\bm\sigma\cdot\left(\bm{n}-\frac{\bm{p}}{\varepsilon_p+m}\right)\chi_{p,\sigma}.
\end{equation}

Analogously as in the positive-energy case, it is clear that
\begin{align}
P_p^{(p)\,\mu}(\phi)&=m\frac{\bar{V}_{p,\sigma}(x)\gamma^{\mu}V_{p,\sigma}(x)}{\bar{V}_{p,\sigma}(x)V_{p,\sigma}(x)}=\pi_p^{(p)\,\mu}(\phi),\\
S_{p,\sigma}^{(p)\,\mu}(\phi)&=-\frac{\bar{V}_{p,\sigma}(x)\gamma^5\gamma^{\mu}V_{p,\sigma}(x)}{\bar{V}_{p,\sigma}(x)V_{p,\sigma}(x)}=\left(\frac{\bm{S}^{(p)}_{p,\sigma,0}(\phi)\cdot\bm{\pi}_p^{(p)}(\phi)}{m},\bm{S}^{(p)}_{p,\sigma,0}(\phi)+\frac{\bm{S}^{(p)}_{p,\sigma,0}(\phi)\cdot\bm{\pi}_p^{(p)}(\phi)}{m[\varepsilon_p^{(p)}(\phi)+m]}\bm{\pi}_p^{(p)}(\phi)\right),
\end{align}
where 
\begin{equation}
\begin{split}
\bm{S}^{(p)}_{p,\sigma,0}(\phi)&=r_{p,\sigma}^{(p)\,\dagger}(\phi)\bm{\sigma}r^{(p)}_{p,\sigma}(\phi)=\bm{s}^{(p)}_{p,\sigma,0}+\frac{\varepsilon_p+m}{p_-}\left(\bm{n}-\frac{\bm{p}}{\varepsilon_p+m}\right)\cdot\bm{s}^{(p)}_{p,\sigma,0}\left(\bm{n}-\frac{\bm{p}}{\varepsilon_p+m}\right)\\
&\quad-\frac{\varepsilon_p^{(p)}(\phi)+m}{p_-}\left[\bm{n}-\frac{\bm\pi_p^{(p)}(\phi)}{\varepsilon_p^{(p)}(\phi)+m}\right]\cdot\bm{s}^{(p)}_{p,\sigma,0}\left[\bm{n}-\frac{\bm\pi_p^{(p)}(\phi)}{\varepsilon_p^{(p)}(\phi)+m}\right]\\
&\quad+\frac{(\varepsilon_p+m)[\varepsilon_p^{(p)}(\phi)+m]}{p^2_-}\left(\bm{n}-\frac{\bm{p}}{\varepsilon_p+m}\right)\cdot\bm{s}^{(p)}_{p,\sigma,0}\\
&\qquad\times\left[\bm{n}-\frac{\bm\pi_p^{(p)}(\phi)}{\varepsilon_p^{(p)}(\phi)+m}\right]\times\left\{\left[\bm{n}-\frac{\bm\pi_p^{(p)}(\phi)}{\varepsilon_p^{(p)}(\phi)+m}\right]\times\left(\bm{n}-\frac{\bm{p}}{\varepsilon_p+m}\right)\right\}.
\end{split}
\end{equation}
Here, we have introduced the pseudovector $\bm{s}^{(p)}_{p,\sigma,0}=\chi_{p,\sigma}^\dagger \bm{\sigma}\chi_{p,\sigma}=-\sigma \bm{s}_p$ (note that with the above choice of $\xi_{p,\sigma}$ and $\chi_{p,\sigma}$, it is $\chi_{p,+}=\xi_{p,-}$ and $\chi_{p,-}=-\xi_{p,+}$).

\section{Fully quasiclassical form of the Volkov propagator}
In this section we derive the Volkov propagator $G(x_1,x_2)$ in the fully quasiclassical form, i.e., the Volkov propagator directly constructed from the fully quasiclassical Volkov states.

We recall that the propagator $G(x_1,x_2)$ is defined via the equation
\begin{equation}
\label{prop1}
i G(x_1,x_2)=\expval{\mathcal{T}\{\Psi(x_1)\bar\Psi(x_2)\}}{0},
\end{equation}
where $\mathcal{T}$ is the time-ordering operator, $\Psi(x)$ is the electron-positron field quantized within the Furry picture, and where $\ket 0$ indicates the vacuum state. The standard representation of the Volkov propagator reads \cite{Ritus_1985} (see also Ref. \cite{Di_Piazza_2018_d} for an expression of the propagator in terms of special functions where the integral over the four-momentum is taken explicitly)
\begin{equation}
\label{G_V}
G(x_1,x_2)=\int\frac{d^4p}{(2\pi)^4}e^{i[S_p(x_1)-S_p(x_2)]}\left[1+e\frac{\hat{n}\hat{A}(\phi_1)}{2p_-}\right]\frac{\hat{p}+m}{p^2-m^2+i0}\left[1-e\frac{\hat{n}\hat{A}(\phi_2)}{2p_-}\right],
\end{equation}
with $\phi_1=(nx_1)$ and $\phi_2=(nx_2)$.

Now, we recall that within the Furry picture the electron-positron field can be expanded in terms of the Volkov states as
\begin{equation}
\Psi(x)=\sum_{\sigma}\int\frac{d^3\bm{p}}{(2\pi)^3}\frac{1}{\sqrt{2\varepsilon_p}}\left[c_{p,\sigma}U_{p,\sigma}(x)+d^{\dag}_{p,\sigma}V_{p,\sigma}(x)\right],
\end{equation}
where $c_{p,\sigma}$ ($d^{\dag}_{p,\sigma}$) are the annihilation (creation) operators of electrons (positrons). By substituting this expression in Eq. (\ref{prop1}) and by recalling the standard anti-commutation rules between the electron and positron creation and annihilation operators, we have that
\begin{equation}
\label{prop2}
\begin{split}
i G(x_1,x_2)&=\theta(x_1^0-x_2^0)\sum_{\sigma}\int\frac{d^3\bm{p}}{(2\pi)^3}\frac{1}{2\varepsilon_p}U_{p,\sigma}(x_1)\bar{U}_{p,\sigma}(x_2)\\
&\quad-\theta(x_2^0-x_1^0)\sum_{\sigma}\int\frac{d^3\bm{p}}{(2\pi)^3}\frac{1}{2\varepsilon_p}V_{p,\sigma}(x_1)\bar{V}_{p,\sigma}(x_2).
\end{split}
\end{equation}
Now, we substitute the fully semiclassical form of the Volkov states in this equation and, after performing the sums over $\sigma$, we obtain
\begin{equation}
\label{prop3}
\begin{split}
&i G(x_1,x_2)=\theta(x_1^0-x_2^0)\int\frac{d^3\bm{p}}{(2\pi)^3}\frac{1}{2\varepsilon_p}\frac{e^{i[S_p(x_1)-S_p(x_2)]}}{\varepsilon_p+m}\\
&\times\begin{pmatrix}
[\varepsilon_p^{(e)}(\phi_1)+m]\mathcal{E}_p^{(e)}(\phi_1)\mathcal{E}_p^{(e)\,\dag}(\phi_2)[\varepsilon_p^{(e)}(\phi_2)+m] &-[\varepsilon_p^{(e)}(\phi_1)+m]\mathcal{E}_p^{(e)}(\phi_1)\mathcal{E}_p^{(e)\,\dag}(\phi_2)\bm{\sigma}\cdot\bm\pi_p^{(e)}(\phi_2)\\
\bm{\sigma}\cdot\bm\pi_p^{(e)}(\phi_1)\mathcal{E}_p^{(e)}(\phi_1)\mathcal{E}_p^{(e)\,\dag}(\phi_2)[\varepsilon_p^{(e)}(\phi_2)+m] & -\bm{\sigma}\cdot\bm\pi_p^{(e)}(\phi_1)\mathcal{E}_p^{(e)}(\phi_1)\mathcal{E}_p^{(e)\,\dag}(\phi_2)\bm{\sigma}\cdot\bm\pi_p^{(e)}(\phi_2)
\end{pmatrix}\\
&-\theta(x_2^0-x_1^0)\int\frac{d^3\bm{p}}{(2\pi)^3}\frac{1}{2\varepsilon_p}\frac{e^{i[S_{-p}(x_1)-S_{-p}(x_2)]}}{\varepsilon_p+m}\\
&\times\begin{pmatrix}
\bm{\sigma}\cdot\bm\pi_p^{(p)}(\phi_1)\mathcal{E}_p^{(p)}(\phi_1)\mathcal{E}_p^{(p)\,\dag}(\phi_2)\bm{\sigma}\cdot\bm\pi_p^{(p)}(\phi_2) &-\bm{\sigma}\cdot\bm\pi_p^{(p)}(\phi_1)\mathcal{E}_p^{(p)}(\phi_1)\mathcal{E}_p^{(p)\,\dag}(\phi_2)[\varepsilon_p^{(p)}(\phi_2)+m]\\
[\varepsilon_p^{(p)}(\phi_1)+m]\mathcal{E}_p^{(p)}(\phi_1)\mathcal{E}_p^{(p)\,\dag}(\phi_2)\bm{\sigma}\cdot\bm\pi_p^{(p)}(\phi_2) & -[\varepsilon_p^{(p)}(\phi_1)+m]\mathcal{E}_p^{(p)}(\phi_1)\mathcal{E}_p^{(p)\,\dag}(\phi_2)[\varepsilon_p^{(p)}(\phi_2)+m]
\end{pmatrix},
\end{split}
\end{equation}
where we have introduced the 2-by-2 matrices
\begin{equation}
\label{dressing1}
\mathcal{E}_p^{(c)}(\phi)=1-\frac{q^{(c)}}{2p_-}\bm\sigma\cdot\left[\bm{n}-\frac{\bm\pi_p^{(c)}(\phi)}{\varepsilon_p^{(c)}(\phi)+m}\right]\bm\sigma\cdot\bm A(\phi),
\end{equation}
with the upper index $c$ taking the values $e$ and $p$ and with $q^{(e)}=e$ and $q^{(p)}=-e$ (recall that in our notation the quantity $e$ is negative).

Exactly as in vacuum, we would like to transform the three-dimensional momentum integrals in four-dimensional ones by using the identity
\begin{equation}
\theta(x_1^0-x_2^0)\frac{f(\varepsilon_p)}{2\varepsilon_p}e^{-i\varepsilon_p(x^0_1-x^0_2)}+\theta(x_2^0-x_1^0)\frac{f(-\varepsilon_p)}{2\varepsilon_p}e^{i\varepsilon_p(x^0_1-x^0_2)}=i\int \frac{dp^0}{2\pi}e^{-ip^0(x^0_1-x^0_2)}\frac{f(p^0)}{p^2-m^2+i0},
\end{equation}
where $f(p^0)$ is an analytic function, which does not vanish at $p^0=\pm \varepsilon_p$ and which is such that the corresponding integrals over the infinite semicircles with $\text{Im}(p^0)\gtrless 0$ vanish. However, Eq. (\ref{prop3}) is not written in the most convenient form for this aim because of the terms $\varepsilon_p+m$, which would give rise to an additional apparent pole at $p^0=-m$ [we know from the traditional form of the Volkov propagator in Eq. (\ref{G_V}) that the only poles to be circumvented are at $p^0=\pm\varepsilon_p$]. This can be avoided by using the identity [see also Eqs. (\ref{sol3}) and (\ref{sol5})]
\begin{equation}
\mathcal{E}_p^{(c)}(\phi)=\frac{\varepsilon_p+m}{2p_-}\bm\sigma\cdot\left[\bm{n}-\frac{\bm\pi_p^{(c)}(\phi)}{\varepsilon_p^{(c)}(\phi)+m}\right]\bm\sigma\cdot\left(\bm{n}-\frac{\bm{p}}{\varepsilon_p+m}\right),
\end{equation}
which implies that
\begin{equation}
\mathcal{E}_p^{(c)}(\phi_1)\mathcal{E}_p^{(c)\,\dag}(\phi_2)=\frac{\varepsilon_p+m}{2p_-}\bm\sigma\cdot\left[\bm{n}-\frac{\bm\pi_p^{(c)}(\phi_1)}{\varepsilon_p^{(c)}(\phi_1)+m}\right]\bm\sigma\cdot\left[\bm{n}-\frac{\bm\pi_p^{(c)}(\phi_2)}{\varepsilon_p^{(c)}(\phi_2)+m}\right].
\end{equation}
This equation allows one to write the propagator in the more convenient form
\begin{equation}
\label{prop4}
\begin{split}
i G(x_1,x_2)&=\theta(x_1^0-x_2^0)\int\frac{d^3\bm{p}}{(2\pi)^3}\frac{1}{2\varepsilon_p}\frac{e^{i[S_p(x_1)-S_p(x_2)]}}{2p_-}\\
&\qquad\times\begin{pmatrix}
\bm{\sigma}\cdot\bm{\Pi}_{+,p}^{(e)}(\phi_1)\bm{\sigma}\cdot\bm{\Pi}_{+,p}^{(e)}(\phi_2) &\bm{\sigma}\cdot\bm{\Pi}_{+,p}^{(e)}(\phi_1)\bm{\sigma}\cdot\bm{n}\bm{\sigma}\cdot\bm{\Pi}_{-,p}^{(e)}(\phi_2)\\
-\bm{\sigma}\cdot\bm{\Pi}_{-,p}^{(e)}(\phi_1)\bm{\sigma}\cdot\bm{n}\bm{\sigma}\cdot\bm{\Pi}_{+,p}^{(e)}(\phi_2) & -\bm{\sigma}\cdot\bm{\Pi}_{-,p}^{(e)}(\phi_1)\bm{\sigma}\cdot\bm{\Pi}_{-,p}^{(e)}(\phi_2)
\end{pmatrix}\\
&\quad-\theta(x_2^0-x_1^0)\int\frac{d^3\bm{p}}{(2\pi)^3}\frac{1}{2\varepsilon_p}\frac{e^{i[S_{-p}(x_1)-S_{-p}(x_2)]}}{2p_-}\\
&\qquad\times\begin{pmatrix}
\bm{\sigma}\cdot\bm{\Pi}_{-,p}^{(p)}(\phi_1)\bm{\sigma}\cdot\bm{\Pi}_{-,p}^{(p)}(\phi_2) &\bm{\sigma}\cdot\bm{\Pi}_{-,p}^{(p)}(\phi_1)\bm{\sigma}\cdot\bm{n}\bm{\sigma}\cdot\bm{\Pi}_{+,p}^{(p)}(\phi_2)\\
-\bm{\sigma}\cdot\bm{\Pi}_{+,p}^{(p)}(\phi_1)\bm{\sigma}\cdot\bm{n}\bm{\sigma}\cdot\bm{\Pi}_{-,p}^{(p)}(\phi_2) & -\bm{\sigma}\cdot\bm{\Pi}_{+,p}^{(p)}(\phi_1)\bm{\sigma}\cdot\bm{\Pi}_{+,p}^{(p)}(\phi_2)
\end{pmatrix},
\end{split}
\end{equation}
where
\begin{equation}
\bm{\Pi}_{\pm,p}^{(c)}(\phi)=[\varepsilon_p^{(c)}(\phi)\pm m]\left[\frac{\bm\pi_p^{(c)}(\phi)}{\varepsilon_p^{(c)}(\phi)\pm m}-\bm{n}\right]=\bm\pi_p^{(c)}(\phi)-[\varepsilon_p^{(c)}(\phi)\pm m]\bm{n}.
\end{equation}
The matrix structure of the above expression of the propagator can be further simplified by using the properties of the Pauli matrices
\begin{align}
\bm{\sigma}\cdot\bm{V}_1\bm{\sigma}\cdot\bm{V}_2&=\bm{V}_1\cdot\bm{V}_2+i\bm{\sigma}\cdot(\bm{V}_1\times\bm{V}_2),\\
\bm{\sigma}\cdot\bm{V}_1\bm{\sigma}\cdot\bm{V}_2\bm{\sigma}\cdot\bm{V}_3&=-i\bm{V}_2\cdot(\bm{V}_1\times\bm{V}_3)+\bm{\sigma}\cdot[(\bm{V}_2\cdot\bm{V}_3)\bm{V}_1-(\bm{V}_1\cdot\bm{V}_3)\bm{V}_2+(\bm{V}_1\cdot\bm{V}_2)\bm{V}_3]
\end{align}
for arbitrary vectors $\bm{V}_1$, $\bm{V}_2$, and $\bm{V}_3$. Then, we obtain
\begin{align}
\begin{split}
&\bm{\sigma}\cdot\bm{\Pi}_{\pm,p}^{(c)}(\phi_1)\bm{\sigma}\cdot\bm{\Pi}_{\pm,p}^{(c)}(\phi_2)=\bm{\Pi}_{\pm,p}^{(c)}(\phi_1)\cdot\bm{\Pi}_{\pm,p}^{(c)}(\phi_2)+i\bm{\sigma}\cdot[\bm{\Pi}_{\pm,p}^{(c)}(\phi_1)\times \bm{\Pi}_{\pm,p}^{(c)}(\phi_2)]\\
&\quad=\frac{[\pi^{(c)}_p(\phi_1)-\pi^{(c)}_p(\phi_2)]^2}{2}+p_-[\varepsilon^{(c)}_p(\phi_1)+\varepsilon^{(c)}_p(\phi_2)\pm 2m]\\
&\qquad+i\bm{\sigma}\cdot\langle\bm{n}\times\{[\varepsilon^{(c)}_p(\phi_2)\pm m]\bm{\pi}^{(c)}_p(\phi_1)-[\varepsilon^{(c)}_p(\phi_1)\pm m]\bm{\pi}^{(c)}_p(\phi_2)\}+\bm{\pi}^{(c)}_p(\phi_1)\times\bm{\pi}^{(c)}_p(\phi_2)\rangle,
\end{split}\\
\begin{split}
&\bm{\sigma}\cdot\bm{\Pi}_{\pm,p}^{(c)}(\phi_1)\bm{\sigma}\cdot\bm{n}\bm{\sigma}\cdot\bm{\Pi}_{\mp,p}^{(c)}(\phi_2)=-i\bm{n}\cdot[\bm{\Pi}^{(c)}_{\pm,p}(\phi_1)\times\bm{\Pi}^{(c)}_{\mp,p}(\phi_2)]\\
&\qquad+\bm{\sigma}\cdot\{[\bm{n}\cdot\bm{\Pi}_{\mp,p}^{(c)}(\phi_2)]\bm{\Pi}_{\pm,p}^{(c)}(\phi_1)-[\bm{\Pi}_{\pm,p}^{(c)}(\phi_1)\cdot\bm{\Pi}_{\mp,p}^{(c)}(\phi_2)]\bm{n}+[\bm{n}\cdot\bm{\Pi}_{\pm,p}^{(c)}(\phi_1)]\bm{\Pi}_{\mp,p}^{(c)}(\phi_2)\}\\
&\quad=-i\bm{n}\cdot[\bm{\pi}^{(c)}_p(\phi_1)\times\bm{\pi}^{(c)}_p(\phi_2)]+\bm{\sigma}\cdot\langle\pm m[\bm{\pi}^{(c)}_p(\phi_1)-\bm{\pi}^{(c)}_p(\phi_2)]-p_-[\bm{\pi}^{(c)}_p(\phi_1)+\bm{\pi}^{(c)}_p(\phi_2)]\\
&\qquad+\{(p_-\mp m)[\varepsilon^{(c)}_p(\phi_1)\pm m]+(p_-\pm m)[\varepsilon^{(c)}_p(\phi_2)\mp m]+(\pi^{(c)}_p(\phi_1)\pi^{(c)}_p(\phi_2))\\
&\qquad-p_-[\varepsilon^{(c)}_p(\phi_1)+\varepsilon^{(c)}_p(\phi_2)]+m^2\}\bm{n}\rangle\\
&\quad=-i\bm{n}\cdot[\bm{\pi}^{(c)}_p(\phi_1)\times\bm{\pi}^{(c)}_p(\phi_2)]+\bm{\sigma}\cdot\Bigg\langle\pm m[\bm{\pi}^{(c)}_p(\phi_1)-\bm{\pi}^{(c)}_p(\phi_2)]-p_-[\bm{\pi}^{(c)}_p(\phi_1)+\bm{\pi}^{(c)}_p(\phi_2)]\\
&\qquad+\left.\left\{\pm m[\varepsilon^{(c)}_p(\phi_2)-\varepsilon^{(c)}_p(\phi_1)]-\frac{[\pi^{(c)}_p(\phi_1)-\pi^{(c)}_p(\phi_2)]^2}{2}\right\}\bm{n}\right\rangle.
\end{split}
\end{align}

Now, it is easy to verify that Eq. (\ref{prop4}) is suitable to introduce the additional integral in $p^0$ by also noticing that by changing in the first ``electron'' matrix $p^{\mu}$ to $-p^{\mu}$, one obtains the second ``positron'' matrix [note that $\pi_{-p}^{(e)\,\mu}(\phi)=-\pi_p^{(p)\,\mu}(\phi)$ and then that $\bm{\Pi}_{+,-p}^{(e)}(\phi)=-\bm{\Pi}_{-,p}^{(p)}(\phi)$]. Thus, we conclude that the propagator can be written as
\begin{equation}
\label{prop5}
G(x_1,x_2)=\int\frac{d^4p}{(2\pi)^4}\frac{e^{i[S_p(x_1)-S_p(x_2)]}}{p^2-m^2+i0}\begin{pmatrix}
G_{UL}(\phi_1,\phi_2) & G_{UR}(\phi_1,\phi_2)\\
G_{BL}(\phi_1,\phi_2) & G_{BR}(\phi_1,\phi_2)
\end{pmatrix},
\end{equation}
where
\begin{align}
G_{UL}(\phi_1,\phi_2)&=m+\frac{(\pi_1-\pi_2)^2}{4p_-}+\frac{\varepsilon_1+\varepsilon_2}{2}+\frac{i\bm{\sigma}}{2p_-}\cdot\{\bm{n}\times[m(\bm{\pi}_1-\bm{\pi}_2)+\varepsilon_2\bm{\pi}_1-\varepsilon_1\bm{\pi}_2]+\bm{\pi}_1\times\bm{\pi}_2\},\\
\begin{split}
G_{BL}(\phi_1,\phi_2)&=\frac{i}{2p_-}\bm{n}\cdot(\bm{\pi}_1\times\bm{\pi}_2)\\
&\quad+\frac{\bm{\sigma}}{2p_-}\cdot\left\{m(\bm{\pi}_1-\bm{\pi}_2)+p_-(\bm{\pi}_1+\bm{\pi}_2)+\left[m(\varepsilon_2-\varepsilon_1)+\frac{(\pi_1-\pi_2)^2}{2}\right]\bm{n}\right\},
\end{split}\\
\begin{split}
G_{UR}(\phi_1,\phi_2)&=-\frac{i}{2p_-}\bm{n}\cdot(\bm{\pi}_1\times\bm{\pi}_2)\\
&\quad+\frac{\bm{\sigma}}{2p_-}\cdot\left\{m(\bm{\pi}_1-\bm{\pi}_2)-p_-(\bm{\pi}_1+\bm{\pi}_2)+\left[m(\varepsilon_2-\varepsilon_1)-\frac{(\pi_1-\pi_2)^2}{2}\right]\bm{n}\right\},
\end{split}\\
G_{BR}(\phi_1,\phi_2)&=m-\frac{(\pi_1-\pi_2)^2}{4p_-}-\frac{\varepsilon_1+\varepsilon_2}{2}+\frac{i\bm{\sigma}}{2p_-}\cdot\{\bm{n}\times[m(\bm{\pi}_1-\bm{\pi}_2)+\varepsilon_1\bm{\pi}_2-\varepsilon_2\bm{\pi}_1]-\bm{\pi}_1\times\bm{\pi}_2\}.
\end{align}
Here, we have introduced the off-shell four-vector
\begin{equation}
\pi_p^{\mu}(\phi)=(\varepsilon_p(\phi),\bm{\pi}_p(\phi))=p^\mu-eA^\mu(\phi)+\frac{e(pA(\phi))}{p_-}n^{\mu}-\frac{e^2A^2(\phi)}{2p_-}n^{\mu},
\end{equation}
with $p_-=p^0-\bm{n}\cdot\bm{p}$ and the short notation $\pi^{\mu}_a=(\varepsilon_a,\bm{\pi}_a)=\pi^{\mu}_p(\phi_a)=(\varepsilon_p(\phi_a),\bm{\pi}_p(\phi_a))$, with $a=1,2$.  

It is interesting to note that the matrices $G_{BL}(\phi_1,\phi_2)$ and $G_{BR}(\phi_1,\phi_2)$ can be obtained by changing $m$ into $-m$ in the matrices $-G_{UR}(\phi_1,\phi_2)$ and $-G_{UL}(\phi_1,\phi_2)$, respectively, such that we only have to work with two independent 2-by-2 matrices.

As a first check, one can easily prove that the propagator $G(x_1,x_2)$ in Eq. (\ref{prop5}) reduces to the free one in the case $A^{\mu}(\phi)=0$. Also, one can explicitly prove that Eq. (\ref{prop5}) is equivalent to Eq. (\ref{G_V}). By applying the identity in Eq. (\ref{p_pi}) and its Dirac conjugated, we have that
\begin{equation}
\begin{split}
&\left[1+e\frac{\hat{n}\hat{A}(\phi_1)}{2p_-}\right](\hat{p}+m)\left[1-e\frac{\hat{n}\hat{A}(\phi_2)}{2p_-}\right]\\
&\quad=\left[1+e\frac{\hat{n}\hat{A}(\phi_1)}{2p_-}\right]\frac{\hat{p}+\hat{p}}{2}\left[1-e\frac{\hat{n}\hat{A}(\phi_2)}{2p_-}\right]+m\left\{1+e\frac{\hat{n}[\hat{A}(\phi_1)-\hat{A}(\phi_2)]}{2p_-}\right\}\\
&\quad=\frac{\hat{\pi}_p(\phi_1)}{2}\left\{1+e\frac{\hat{n}[\hat{A}(\phi_1)-\hat{A}(\phi_2)]}{2p_-}\right\}+\left\{1+e\frac{\hat{n}[\hat{A}(\phi_1)-\hat{A}(\phi_2)]}{2p_-}\right\}\frac{\hat{\pi}_p(\phi_2)}{2}\\
&\qquad+m\left\{1+e\frac{\hat{n}[\hat{A}(\phi_1)-\hat{A}(\phi_2)]}{2p_-}\right\}.
\end{split}
\end{equation}
Now, we use the identity $e\hat{n}[\hat{A}(\phi_1)-\hat{A}(\phi_2)]=-\hat{n}(\hat{\pi}_1-\hat{\pi}_2)$ and we obtain (see Refs. \cite{Hartin_2016,Adamo_2019} for equivalent expressions of the Volkov propagator, with the matrix integrand depending only on the electron dressed kinetic four-momentum)
\begin{equation}
\label{G_V_inte}
\begin{split}
\left[1+e\frac{\hat{n}\hat{A}(\phi_1)}{2p_-}\right](\hat{p}+m)\left[1-e\frac{\hat{n}\hat{A}(\phi_2)}{2p_-}\right]&=\frac{\hat{\pi}_1+\hat{\pi}_2}{2}+\frac{(\pi_1-\pi_2)^2}{4p_-}\hat{n}\\
&\quad+\frac{1}{4p_-}(\hat{n}\hat{\pi}_2\hat{\pi}_1-\hat{\pi}_1\hat{\pi}_2\hat{n})+m\left[1-\frac{\hat{n}(\hat{\pi}_1-\hat{\pi}_2)}{2p_-}\right].
\end{split}
\end{equation}
Finally, by using the standard representation of the Dirac matrices, one can show that the 4-by-4 matrix in Eq. (\ref{G_V_inte}) can be written in blocks of 2-by-2 matrices and that it coincides with that in Eq. (\ref{prop5}). Observe that the above checking procedure of the equivalence of the standard form and the quasiclassical form of the Volkov propagator can also be employed as an alternative derivation of Eq. (\ref{prop5}).

\section{The one-loop tadpole contribution in an arbitrary plane wave}

\begin{figure}
\begin{center}
\includegraphics[width=0.5\columnwidth]{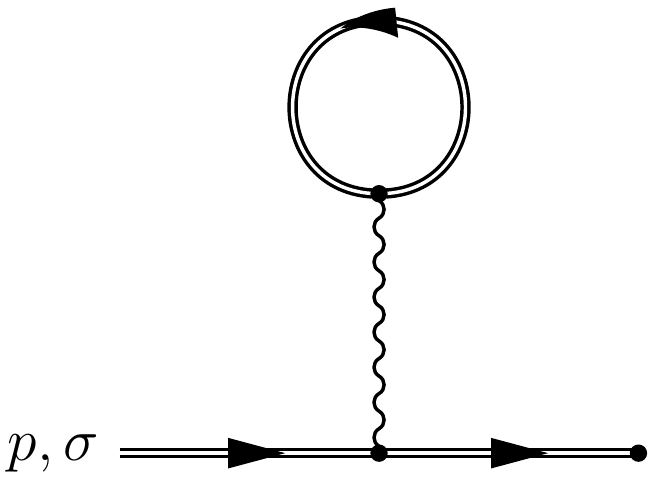}
\caption{The one-loop tadpole contribution to the Volkov state $U_{p,\sigma}(x)$.}
\label{TP}
\end{center}
\end{figure}

The one-loop tadpole contribution to the Volkov state $U_{p,\sigma}(x)$ is represented in Fig. \ref{TP}, where all double lines indicate either the Volkov state or the Volkov propagator.
By indicating the corresponding amplitude as $\delta U^{(1)}_{p,\sigma}(x)$ and by applying the usual Feynman rules, one obtains
\begin{equation}
\label{delta_U}
\begin{split}
\delta U^{(1)}_{p,\sigma}(x)&=\int d^4yd^4z\,iG(x,y)(-ie\gamma^{\mu})U_{p,\sigma}(y)(-i)D_{\mu\nu}(y-z)(-1)\text{tr}[(-ie\gamma^{\nu})iG(z,z)]\\
&=ie^2\int d^4yd^4z\,G(x,y)\gamma^{\mu}U_{p,\sigma}(y)D_{\mu\nu}(y-z)\text{tr}[\gamma^{\nu}G(z,z)],
\end{split}
\end{equation}
where $D_{\mu\nu}(x-y)$ is the photon propagator in the Feynman gauge
\begin{equation}
\label{D}
D^{\mu\nu}(x-y)=\int\frac{d^4k}{(2\pi)^4}\frac{\eta^{\mu\nu}}{k^2+i0}e^{-i(k(x-y))}.
\end{equation}

Following Schwinger \cite{Schwinger_1951}, we recall that the quantity $G(x,x)$ has to be meant as 
\begin{equation}
\label{G_xx}
G(x,x)=\frac{1}{2}\left[\lim_{t_y\to t_x^+}G(t_x,\bm{x},t_y,\bm{x})+\lim_{t_y\to t_x^-}G(t_x,\bm{x},t_y,\bm{x})\right],
\end{equation}
i.e., the first (second) limit has to be taken with $t_y>t_x$ ($t_y<t_x$). Now, starting from the general definition in Eq. (\ref{prop2}), we have that
\begin{equation}
G(x,x)=-\frac{i}{2}\sum_{\sigma}\int\frac{d^3\bm{p}}{(2\pi)^3}\frac{1}{2\varepsilon_p}U_{p,\sigma}(x)\bar{U}_{p,\sigma}(x)+\frac{i}{2}\sum_{\sigma}\int\frac{d^3\bm{p}}{(2\pi)^3}\frac{1}{2\varepsilon_p}V_{p,\sigma}(x)\bar{V}_{p,\sigma}(x).
\end{equation}
We recall that the trace $ie\text{tr}[\gamma^{\mu}G(x,x)]$, which corresponds to the tadpole part of the diagram in Fig. \ref{TP}, coincides with the vacuum four-current density $J_v^{\mu}(x)$ \cite{Schwinger_1951}:
\begin{equation}
\begin{split}
J_v^{\mu}(x)&=ie\text{tr}[\gamma^{\mu}G(x,x)]\\
&=\frac{e}{2}\sum_{\sigma}\int\frac{d^3\bm{p}}{(2\pi)^3}\frac{1}{2\varepsilon_p}\bar{U}_{p,\sigma}(x)\gamma^{\mu}U_{p,\sigma}(x)-\frac{e}{2}\sum_{\sigma}\int\frac{d^3\bm{p}}{(2\pi)^3}\frac{1}{2\varepsilon_p}\bar{V}_{p,\sigma}(x)\gamma^{\mu}V_{p,\sigma}(x).
\end{split}
\end{equation}
This equation transparently relates the vacuum four-current density with the electron and positron four-current densities proportional to $\bar{U}_{p,\sigma}(x)\gamma^{\mu}U_{p,\sigma}(x)=2\pi_p^{(e)\,\mu}(\phi)$ and to $\bar{V}_{p,\sigma}(x)\gamma^{\mu}V_{p,\sigma}(x)=2\pi_p^{(p)\,\mu}(\phi)$, respectively (see also \cite{Landau_b_4_1982}). By using these identities, we obtain that
\begin{equation}
\label{J_v}
J_v^{\mu}(x)=e\int\frac{d^3\bm{p}}{(2\pi)^3}\frac{1}{\varepsilon_p}[\pi_p^{(e)\,\mu}(\phi)-\pi_p^{(p)\,\mu}(\phi)]=-2e^2\int\frac{d^3\bm{p}}{(2\pi)^3}\frac{1}{\varepsilon_p}\left[A^{\mu}(\phi)-\frac{(pA(\phi))}{p_-}n^{\mu}\right].
\end{equation}
This expression, although divergent, is manifestly gauge invariant because it is the difference between the electron and the positron kinetic four-momenta in the plane wave. Also, it shows that only linear terms in the plane wave electromagnetic field contribute to the tadpole (see also Ref. \cite{Ahmadiniaz_2019}).

The above expression of the vacuum four-current can also be obtained directly by starting from the expression of the propagator $G(x_1,x_2)$ in Eq. (\ref{G_V}). Indeed, we have that
\begin{equation}
\begin{split}
J_v^{\mu}(x)&=ie\int\frac{d^4p}{(2\pi)^4}\text{tr}\left\{\gamma^{\mu}\left[1+e\frac{\hat{n}\hat{A}(\phi)}{2p_-}\right]\frac{\hat{p}+m}{p^2-m^2+i0}\left[1-e\frac{\hat{n}\hat{A}(\phi)}{2p_-}\right]\right\}\\
&=ie\int\frac{d^4p}{(2\pi)^4}\frac{1}{p^2-m^2+i0}\text{tr}\left\{\gamma^{\mu}\left[1+e\frac{\hat{n}\hat{A}(\phi)}{2p_-}\right]\hat{p}\left[1-e\frac{\hat{n}\hat{A}(\phi)}{2p_-}\right]\right\},
\end{split}
\end{equation}
where we used the fact that the trace of an odd number of gamma matrices vanishes and Eq. (\ref{p_pi}). By exploiting the symmetry properties of the integrand, we have
\begin{equation}
\label{J}
J_v^{\mu}(x)=-4ie^2\int\frac{d^4p}{(2\pi)^4}\frac{1}{p^2-m^2+i0}\left[A^{\mu}(\phi)-\frac{(pA(\phi))}{p_-}n^{\mu}\right],
\end{equation}
and then the above presented expression of the vacuum four-current is obtained by performing the integral over $p^0$ via the residue method
\begin{equation}
\int\frac{dp^0}{2\pi}\frac{1}{(p^0)^2-\varepsilon_p^2+i0}=-\frac{i}{2\varepsilon_p},
\end{equation}
this result being independent on whether one closes the path on the infinite semicircle with $\text{Im}(p^0)>0$ or the one with $\text{Im}(p^0)<0$.

\begin{figure}
\begin{center}
\includegraphics[width=0.5\columnwidth]{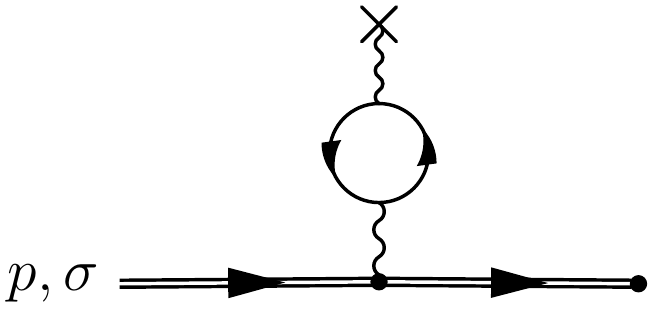}
\caption{The perturbative representation of the one-loop tadpole contribution to the Volkov state $U_{p,\sigma}(x)$. The vertex with the cross corresponds to the external plane wave.}
\label{TP_Pert}
\end{center}
\end{figure}

Since the vacuum four-current is linear in the external field, it must be possible to obtain its expression starting from the Feynman diagram in Fig. \ref{TP_Pert}.
It is useful to derive the vacuum four-current via the perturbative approach because it will give a hint on how to renormalize its (divergent) expression, which is what we are going to do below (see Ref. \cite{Brouder_2002} for a theory of renormalization of QED in an external field, which would lead to the same result as the one obtained below). By applying Feynman rules in vacuum, we have that
\begin{equation}
\label{J_Pert}
iJ_v^{\mu}(x)=-\int d^4y\,\text{tr}[-ie\gamma^{\mu}iG_0(x-y)(-ie)\hat{A}(\phi_y)iG_0(y-x)],
\end{equation}
where
\begin{equation}
\label{G_0}
G_0(x-y)=\int\frac{d^4p}{(2\pi)^4}\frac{\hat{p}+m}{p^2-m^2+i0}e^{-i(p(x-y))},
\end{equation}
is the free electron propagator. By writing the plane-wave four-vector potential as
\begin{equation}
\label{A_F}
A^{\mu}(\phi)=\int\frac{d\omega}{2\pi}e^{-i\omega\phi}\tilde{A}^{\mu}(\omega)=\int\frac{d\omega}{2\pi}e^{-i(kx)}\tilde{A}^{\mu}(\omega),
\end{equation}
where $k^{\mu}=\omega n^{\mu}$, we obtain
\begin{equation}
\begin{split}
J_v^{\mu}(x)&=ie^2\int \frac{d^4p}{(2\pi)^4}\int\frac{d\omega}{2\pi}\frac{e^{-i(kx)}}{p^2-m^2+i0}\frac{\text{tr}[\gamma^{\mu}(\hat{p}+m)\hat{\tilde{A}}(\omega)(\hat{p}-\hat{k}+m)]}{(p-k)^2-m^2+i0}\\
&=-2ie^2\int \frac{d^4p}{(2\pi)^4}\int\frac{d\omega}{2\pi}\frac{e^{-i(kx)}}{(kp)}\left[\frac{1}{p^2-m^2+i0}-\frac{1}{(p-k)^2-m^2+i0}\right]\\
&\quad\times[(p\tilde{A}(\omega))(2p^{\mu}-k^{\mu})-\tilde{A}^{\mu}(\omega)(p^2-m^2-(kp))],\\
&=-2ie^2\int \frac{d^4p}{(2\pi)^4}\int\frac{d\omega}{2\pi}\frac{e^{-i(kx)}}{(kp)}\left[\frac{-(p\tilde{A}(\omega))k^{\mu}+\tilde{A}^{\mu}(\omega)(kp)}{p^2-m^2+i0}-\frac{(p\tilde{A}(\omega))k^{\mu}-\tilde{A}^{\mu}(\omega)(kp)}{p^2-m^2+i0}\right]\\
&=-4ie^2\int \frac{d^4p}{(2\pi)^4}\int\frac{d\omega}{2\pi}\frac{e^{-i(kx)}}{(kp)}\frac{\tilde{A}^{\mu}(\omega)(kp)-(p\tilde{A}(\omega))k^{\mu}}{p^2-m^2+i0},
\end{split}
\end{equation}
where, in the last two steps we have exploited the symmetry properties of the integrand. At this point, the quantity $\omega$ simplifies in the preexponential function and, by using Eq. (\ref{A_F}), we again obtain Eq. (\ref{J}).

Now, the perturbative expression of the vacuum four-current $J_v^{\mu}(x)$ in Eq. (\ref{J_Pert}) gives us a hint on how to renormalize it. In fact, by using the standard definition of the vacuum polarization operator
\begin{equation}
\Pi^{\mu\nu}(x-y)=i\text{tr}[-ie\gamma^{\mu}iG_0(x-y)(-ie)\gamma^{\nu}iG_0(y-x)],
\end{equation}
we obtain
\begin{equation}
J_v^{\mu}(x)=\int d^4y\,\Pi^{\mu\nu}(x-y)A_{\nu}(\phi_y)=\int\frac{d^4q}{(2\pi)^4}\Pi^{\mu\nu}(q)e^{-i(qx)}A_{\nu}(q),
\end{equation}
or $J_v^{\mu}(q)=\Pi^{\mu\nu}(q)A_{\nu}(q)$ [note that $A^{\mu}(q)=\int d^4x e^{i(qx)}A^{\mu}(\phi)=(2\pi)^3\delta(q_-)\delta^2(\bm{q}_{\perp})\tilde{A}^{\mu}(q_+)$, where $q_+=(q^0+\bm{n}\cdot\bm{q})/2$]. Lorentz- and gauge-invariance imply that $\Pi^{\mu\nu}(q)$ has the form $\Pi^{\mu\nu}(q)=(q^2\eta^{\mu\nu}-q^{\mu}q^{\nu})\Pi(q^2)$ and the standard renormalization of the polarization operator amounts to replace the function $\Pi(q^2)$ with the function $\Pi_r(q^2)=\Pi(q^2)-\Pi(0)$, such that $\Pi_r(q^2)$ can be written as $\Pi_r(q^2)=q^2\Phi_r(q^2)$, with $\Phi_r(q^2)$ being finite at $q^2=0$ \cite{Peskin_b_1995}. In this way, we finally obtain that the renormalized vacuum four-current $J_{v,r}^{\mu}(q)=\Pi_r(q^2)(q^2\eta^{\mu\nu}-q^{\mu}q^{\nu})A_{\nu}(q)=q^2\Phi_r(q^2)(q^2\eta^{\mu\nu}-q^{\mu}q^{\nu})(2\pi)^3\delta(q_-)\delta^2(\bm{q}_{\perp})\tilde{A}_{\nu}(q_+)=0$ (see also Ref. \cite{Ahmadiniaz_2019}).

Finally, Eq. (\ref{delta_U}) indicates that in order to compute the correction $\delta U^{(1)}_{p,\sigma}(x)$, we need the quantity
\begin{equation}
B^{\mu}(x)=\int d^4y\,D^{\mu\nu}(x-y)J_{v,\nu}(y),
\end{equation}
which also has to be renormalized. This means to use the renormalized vacuum four-current instead of $J_v^{\mu}(x)$ and then, by passing to momentum space, we obtain for the renormalized four-vector $B_r^{\mu}(q)$:
\begin{equation}
B_r^{\mu}(q)=D^{\mu\nu}(q)J_{v,r,\nu}(q)=\Phi_r(q^2)(q^2\eta^{\mu\nu}-q^{\mu}q^{\nu})(2\pi)^3\delta(q_-)\delta^2(\bm{q}_{\perp})\tilde{A}_{\nu}(q_+)=0,
\end{equation}
where we have used the fact that if $q_-=0$ and $\bm{q}_{\perp}=\bm{0}$, then $q^{\mu}=q_+n^{\mu}$ and $(q\tilde{A}(q_+))=0$. Thus, we conclude that the renormalized correction to the Volkov states as due to the one-loop tadpole diagram identically vanishes in a plane wave after renormalization.

The above situation is somewhat different than that in a constant background electromagnetic field where, although the vacuum four-current vanishes, the tadpole (which includes the photon propagator) does not except for the special case of a constant-crossed field, corresponding to a plane wave with zero frequency \cite{Gies_2017,Karbstein_2017,Edwards_2017,Ahmadiniaz_2017,Ahmadiniaz_2019}. More specifically, although the contribution to the tadpole linear in the background field is also renormalized out in the case of a constant field like in the plane-wave case, higher-order contributions proportional due to Furry theorem to odd powers of the field do not vanish, which is related to the fact that, unlike in a plane wave, the electromagnetic field invariants do not vanish for a constant (non-crossed) field  \cite{Gies_2017,Karbstein_2017,Edwards_2017,Ahmadiniaz_2017,Ahmadiniaz_2019}. As an final technical remark, we also observe that the four-momentum flowing in the photon propagator identically vanishes in a constant field and encountered integrals like $\int d^4k\, \delta^4(k)k^{\mu}k^{\nu}/k^2$ can be shown to be finite (and equal to $\eta^{\mu\nu}/4$) \cite{Gies_2017,Karbstein_2017,Edwards_2017,Ahmadiniaz_2017,Ahmadiniaz_2019}. The situation is different here because the four-momentum entering the polarization operator does not vanish, it is lightlike and it is also orthogonal to the external plane-wave four-vector potential.

\section{Conclusions}
In conclusion, we have first presented an alternative derivation of the fully quasiclassical form of the Volkov states, simpler than the original one in Ref. \cite{Di_Piazza_2021_a}. Then, we have used these states to construct an alternative form of the Volkov propagator, which depends, apart from the actions in the exponential functions, only on the dressed kinetic four-momentum of an electron in a plane wave. This form highlights the properties of the Volkov propagator under a generic gauge transformation of the plane wave and is expressed as four blocks of 2-by-2 matrices. Among these matrices, only two are independent in the sense that the other two can be obtained by a simple substitution rule. 

Due to the easy multiplication rules of the Pauli matrices, the obtained expression of the propagator is conveniently used when performing calculations via the quasiclassical form of the Volkov states. In this respect, the present results complement those in Ref. \cite{Di_Piazza_2021_a} and provide the remaining tool to compute strong-field QED probabilities in a strong plane wave by manipulating pre-exponential functions explicitly depending only on the leptons dressed kinetic four-momenta. Already the relatively straightforward matrix manipulations in the computation of the probabilities of nonlinear Compton scattering and nonlinear Breit-Wheeler pair production in Ref. \cite{Di_Piazza_2021_a}, which are ultimately expressed as traces of two-dimensional matrices, give an idea, although the spin dynamics was ignored there, of the envisaged simplifications in investigating higher-order processes and radiative corrections by means of the quasiclassical Volkov states (and propagator).

Finally, in relation to the electron propagator, we have computed the vacuum four-current density in a plane wave. The related one-loop tadpole contribution to an arbitrary Feynman diagram has then been shown to identically vanish after renormalization. 

\section*{Acknowledgments}
The present article is also supported by the Collaborative Research Centre 1225 funded by Deutsche Forschungsgemeinschaft (DFG, German Research Foundation)-Project-ID 273811115-SFB 1225. ADP gratefully acknowledges insightful discussions with Anton Ilderton and Felix Karbstein.

\section*{Appendix A: Derivation of the quasiclassical spinor}
In this appendix we shall explicitly prove that the two-dimensional spinor in Eq. (\ref{sol2}) satisfies the differential equation (\ref{difeq4}). To see this, we evaluate its derivative with respect to $\phi$. We have (for the sake of notational simplicity we omit the dependence on $\phi$ in some equations below)
\begin{equation}
\label{der1}
\begin{split}
r_{p,\sigma}^{(e)\prime}(\phi)&=\frac{e(\varepsilon_p^{(e)}+m)}{2p_-\sqrt{(\varepsilon_p^{(e)}+m)(\varepsilon_p+m)}}\bigg\{\bm\sigma\cdot\left[\frac{\bm\pi_p^{(e)\prime}}{\varepsilon_p^{(e)}+m}-\frac{\varepsilon_p^{(e)\prime}}{(\varepsilon_p^{(e)}+m)^2}\bm\pi_p^{(e)}\right]\bm\sigma\cdot\bm A\\
&\quad-\bm\sigma\cdot\left(\bm n-\frac{\bm\pi_p^{(e)}}{\varepsilon_p^{(e)}+m}\right)\bm\sigma\cdot\bm A'\bigg\}\xi_{p,\sigma}+
\frac{\varepsilon_p^{(e)\prime}}{2\sqrt{(\varepsilon_p^{(e)}+m)(\varepsilon_p+m)}}r_{p,\sigma}(\phi)\\
&=a\bigg\{\bm\pi_p^{(e)}\cdot\bm E\left[1-\frac{e}{2p_-}\bm\sigma\cdot\left(\bm n-\frac{\bm\pi_p^{(e)}}{\varepsilon_p^{(e)}+m}\right)\bm\sigma\cdot\bm A\right]\\
&\quad+\frac{e}{p_-}\bm\sigma\cdot\left(\varepsilon_p^{(e)}\bm E+\bm\pi_p^{(e)}\times\bm B-\frac{\bm\pi_p^{(e)}\cdot\bm E}{\varepsilon_p^{(e)}+m}\bm\pi_p^{(e)}\right)\bm\sigma\cdot\bm A
-\bm\sigma\cdot[\bm\pi_p^{(e)}-(\varepsilon_p^{(e)}+m)\bm n]\bm\sigma\cdot\bm E\bigg\}\xi_{p,\sigma}\\
&=a\bigg\{-\bm\pi_p^{(e)}\cdot\bm E\left[\frac{e}{2p_-}\bm\sigma\cdot\left(\frac{\bm\pi_p^{(e)}}{\varepsilon_p^{(e)}+m}+\bm n\right)\bm\sigma\cdot\bm A\right]+\frac{e}{p_-}\bm\sigma\cdot\left(\varepsilon_p^{(e)}\bm E+\bm\pi_p^{(e)}\times\bm B\right)\bm\sigma\cdot\bm A\\
&\quad
-i\bm\sigma\cdot[\bm\pi_p^{(e)}-(\varepsilon_p^{(e)}+m)\bm n]\times\bm E\bigg\}\xi_{p,\sigma},
\end{split}
\end{equation}
where
\begin{equation}
a(\phi)=\frac{e}{2p_-\sqrt{[\varepsilon_p^{(e)}(\phi)+m](\varepsilon_p+m)}}
\end{equation}
and where we have also used the identities
\begin{align}
\varepsilon_p^{(e)\prime}(\phi)&=\frac{e}{p_-}\bm\pi_p^{(e)}(\phi)\cdot\bm E(\phi),\\
\bm\pi_p^{(e)\prime}(\phi)&=\frac{e}{p_-}[\varepsilon_p^{(e)}(\phi) \bm E(\phi)+\bm\pi_p^{(e)}(\phi)\times\bm B(\phi)]
\end{align}
corresponding to the Lorentz four-force equation in the plane wave. In the last equality of Eq. (\ref{der1}) we have also used the fact that $\bm\sigma\cdot\bm\pi_p^{(e)}(\phi) \bm\sigma\cdot\bm E(\phi)=\bm\pi_p^{(e)}(\phi)\cdot\bm E(\phi)+i\bm\sigma\cdot[\bm\pi_p^{(e)}(\phi)\times\bm E(\phi)]$ and that $\bm\sigma\cdot\bm n\bm\sigma\cdot\bm E(\phi)=i\bm\sigma\cdot[\bm n\times\bm E(\phi)]$.

For the right-hand side of  Eq. (\ref{difeq4}), which we denote here as $R(\phi)$, one finds that
\begin{equation}
\label{der2}
\begin{split}
R(\phi)&=i a\bm\sigma\cdot\left[(\varepsilon_p^{(e)}+m)\bm B-\bm\pi_p^{(e)}\times\bm E\right]\left[1-\frac{e}{2p_-}\bm\sigma\cdot\left(\bm n-\frac{\bm\pi_p^{(e)}}{\varepsilon_p^{(e)}+m}\right)\bm\sigma\cdot\bm A\right]\xi_{p,\sigma}\\
&=a\bigg\{i(\varepsilon_p^{(e)}+m)\bm\sigma\cdot\bm B-i\bm\sigma\cdot(\bm\pi_p^{(e)}\times\bm E)+\frac{i e}{2p_-}\left[\bm B\cdot\bm\pi_p^{(e)}+\bm n\cdot(\bm\pi_p^{(e)}\times\bm E)\right]\bm{\sigma}\cdot\bm{A}\\
&\quad
+\frac{e}{2p_-}\bm\sigma\cdot\left[(\varepsilon_p^{(e)}+m)\bm E+\frac{(\bm\pi_p^{(e)}\times\bm E)\times\bm\pi_p^{(e)}}{\varepsilon_p^{(e)}+m}-\bm B\times\bm\pi_p^{(e)}-(\bm\pi_p^{(e)}\times\bm E)\times\bm n\right]\bm\sigma\cdot\bm A\bigg\}\xi_{p,\sigma}.
\end{split}
\end{equation}
Given that $\bm B(\phi)=\bm n\times\bm E(\phi)$, the first two terms coincide with the last two terms of Eq. (\ref{der1}). Also, note that $\bm n\cdot[\bm\pi_p^{(e)}(\phi)\times\bm E(\phi)]=\bm\pi_p^{(e)}(\phi)\cdot[\bm E(\phi)\times\bm n]=-\bm\pi_p^{(e)}(\phi)\cdot\bm B(\phi)$ such that the term in the first line proportional to $\bm{\sigma}\cdot\bm{A}(\phi)$ vanishes. All that remains is to examine the terms in the second line of Eq. (\ref{der2}). Starting from the first two, we have that, since $[\bm\pi_p^{(e)}(\phi)\times\bm E(\phi)]\times\bm\pi_p^{(e)}(\phi)=\bm E(\phi)[\varepsilon_p^{(e)\,2}(\phi)-m^2]-\bm\pi_p^{(e)}(\phi)[\bm\pi_p^{(e)}(\phi)\cdot\bm E(\phi)]$, then it is:
\begin{equation}
[\varepsilon_p^{(e)}(\phi)+m]\bm E(\phi)+\frac{[\bm\pi_p^{(e)}(\phi)\times\bm E(\phi)]\times\bm\pi_p^{(e)}(\phi)}{\varepsilon_p^{(e)}(\phi)+m}=2\varepsilon_p^{(e)}(\phi)\bm E(\phi)-\frac{\bm\pi_p^{(e)}(\phi)\cdot\bm E(\phi)}{\varepsilon_p^{(e)}(\phi)+m}\bm\pi_p^{(e)}(\phi).
\end{equation}
By using the known identities for the double cross product, we obtain that $\bm n\times[\bm\pi_p^{(e)}(\phi)\times\bm E(\phi)]=-\bm E(\phi)[\varepsilon_p^{(e)}(\phi)-p_-]$ and then that
\begin{equation}
\label{der3}
\begin{split}
R(\phi)&=a\bigg\{i(\varepsilon_p^{(e)}+m)\bm\sigma\cdot\bm B-i\bm\sigma\cdot(\bm\pi_p^{(e)}\times\bm E)+\frac{e}{p_-}\bm\sigma\cdot\left(\varepsilon_p^{(e)}\bm E-\bm B\times\bm\pi_p^{(e)}\right)\bm\sigma\cdot\bm A\\
&\quad
-\frac{e}{2p_-}\bm\sigma\cdot\left[\bm \pi_p^{(e)}\times\bm B+\bm E(\varepsilon_p^{(e)}-p_-)+\frac{\bm\pi_p^{(e)}\cdot\bm E}{\varepsilon_p^{(e)}+m}\bm\pi_p^{(e)}\right]\bm\sigma\cdot\bm A\bigg\}\xi_{p,\sigma}.
\end{split}
\end{equation}
Finally, since $\bm\pi_p^{(e)}(\phi)\times[\bm n\times\bm E(\phi)]=\bm n [\bm\pi_p^{(e)}(\phi)\cdot\bm E(\phi)]-\bm E(\phi)[\varepsilon_p^{(e)}(\phi)-p_-]$, we conclude that the quantities $R(\phi)$ and $r_{p,\sigma}^{(e)\prime}(\phi)$ indeed coincide.

%


\end{document}